\documentclass[prb,twocolumn]{revtex4}

\usepackage{graphicx}
\usepackage{dcolumn}
\usepackage{amsmath}
\usepackage{epsfig}
\def\sbf{$\beta$-(BDA-TTP)$_2$SbF$_6$}
\def\bda{$\beta$-(BDA-TTP)$_{2}X$}
\def\ibr{$\beta$-(BEDT-TTF)$_2$IBr$_2$}
\def\betai{$\beta$-(BEDT-TTF)$_2$I$_3$}
\def\betah{$\beta_H$-(BEDT-TTF)$_2$I$_3$}
\def\kappai{$\kappa$-(BEDT-TTF)$_2$I$_3$}
\def\cuncs{$\kappa$-(BEDT-TTF)$_2$Cu(NCS)$_2$}
\def\cunbr{$\kappa$-(BEDT-TTF)$_2$Cu[N(CN)$_2$]Br}
\def\nh4{$\alpha$-(BEDT-TTF)$_2$NH$_4$Hg(SCN)$_4$}
\def\gacl4{$\lambda$-(BETS)$_2$GaCl$_4$}
\def\sf5{$\beta^{\prime\prime}$-(BEDT-TTF)$_2$SF$_5$CH$_2$CF$_2$SO$_3$}

\def\delc{$\Delta C_e/\gamma T_c$}
\def\lph{$\lambda_{EP}$}

\begin {document}

\title[Short Title]{Fermiology and superconductivity studies on the non-tetrachalcogenafulvalene structured
organic superconductor \sbf}

\author{E. S. Choi, E. Jobilong, A. Wade, E. Goetz, J. S. Brooks}
\affiliation{National High Magnetic Field Laboratory, Florida
State University, Tallahassee, Florida 32310}
\author{J. Yamada}
\affiliation{Department of Material Science, Faculty of Science,
Himeji Institute of Technology, Hyogo 678-1297, Japan,}
\author{T. Mizutani, T. Kinoshita, M. Tokumoto}
\affiliation{Nanotechnology Research Institute, Natl Inst. of
Advanced Industrial Science and Technology, Ibaraki 305-8568,
Japan}
\begin{abstract}
The quantum oscillatory effect and superconductivity in a
non-tetrachalcogenafulvalene (TCF) structure based organic
superconductor \sbf~ are studied. Here the  Shubnikov-de Haas
effect (SdH) and angular dependent magnetoresistance oscillations
(AMRO) are observed. The oscillation frequency associated with a
cylindrical Fermi surface is found to be about 4050 tesla, which
is also verified by the tunnel diode oscillator (TDO) measurement.
The upper critical field ($H_{c2}$) measurement in a tilted
magnetic field and the TDO measurement in the mixed state reveal a
highly anisotropic superconducting nature in this material. We
compared physical properties of \sbf~with typical TCF structure
based quasi two-dimensional organic conductors. A notable feature
of \sbf~superconductor is a large value of effective cyclotron
mass $m_c^*$=12.4$\pm$1.1$m_e$, which is the largest yet found in
an organic superconductor. A possible origin of the enhanced
effective mass and its relation to the superconductivity are
briefly discussed.
\end{abstract}

\pacs{74.70.Kn, 71.18.+y}
\maketitle

\section{Introduction}

\begin{figure}[bp]
\epsfig{file=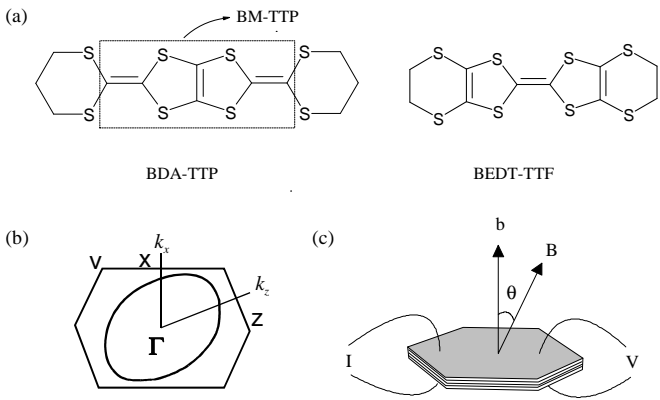, bb=26 460 484 766 clip=,width=\linewidth}
\caption{(a)Donor molecular structure of BDA-TTP and BEDT-TTF. (b)
Theoretical Fermi surface of \sbf. (c) The configuration of
contacts and sample orientations for the magnetoresistance
measurement.}\label{fig1}
\end{figure}

Recent synthesis of new organic superconductors \bda~(where
BDA-TTP is
2,5-bis(1,3-dithian-2-ylidene)-1,3,4,6,-tetrathiapentalene and
$X^-$ = SbF$_6^-$, AsF$_6^-$ and PF$_6^-$) has attracted interest
because the $\pi$-electron donor molecules have a unique
structure.\cite{yamada} Most charge transfer salt organic
superconductors have $\pi$-electron donors with
tetrachalcogenafulvalene (TCF) structures, but BDA-TTP is the
derivative of BM-TTP (2,5-bis(methylene)-1,3,4,6-
tetrathiapentalene) whose structure is different as shown in Fig.
\ref{fig1}(a). Though the $\pi$-electron overlapping along the
donor-stacking directions is known to play an important role in
the superconductivity as well as metallic properties in organic
conductors, the detailed mechanism is not yet well understood
.\cite{ishiguro} Since the electron-molecular-vibration (EMV)
coupling will be different between TCF and BM-TTP structured
molecules, they provide new materials for comparative studies.

The superconductivity of \sbf~ ($T_c \approx$ 6.5 K) was
characterized from the temperature dependence of the upper
critical field $H_{c2}$ and the specific heat measurement by
Shimojo et al.\cite{shimojo} The overall properties were found to
be similar to those of the other layered organic superconductors.
The specific heat jump at superconducting transition is about
$\Delta C_e /\gamma T_c$=1.1, which is smaller than the value 1.43
for weak-coupling BCS superconductor. However, the quantum
oscillatory effects, which are powerful tools for determining the
electronic structure of organic conductors, have not previously
been reported. According to a tight binding band calculation
result, a nearly isotropic cylindrical Fermi surface exists as
shown in Fig. \ref{fig1}(b), which is characteristic to
$\beta$-type organic conductors.\cite{yamada}

In this paper, we report the first observation of Shubnikov-de
Haas (SdH) effect and the angular-dependent magnetoresistance
oscillations (AMRO) of \sbf~ superconductor. We also performed a
tunnel diode oscillator (TDO) measurement on the same material and
observed quantum oscillations corresponding to the cylindrical
Fermi surface. The measured frequency of SdH oscillation is in
good agreement with the band calculation result. Superconducting
properties are studied by the angular dependent $H_{c2}$ and TDO
measurements. Finally, fermiological and superconducting
properties based on this work and previous reports are compared to
other organic superconductors.

\section{Experimental Details}
Single crystal samples were prepared by typical electrochemical
technique.\cite{yamada} For the magnetoresistance measurement, the
four probe method was employed with low frequency lock-in
techniques. Four gold wires were attached to a plate-like samples
using graphite paint, and typically 10 $\mu$A current was applied
perpendicular to the conducting planes ($ac$-plane). The samples
were mounted on a rotator probe of a dilution refrigerator, which
is situated in a 33 tesla resistive magnet at the National High
Magnetic Field Laboratory, Tallahassee. The contact configuration
and the definition of rotation angle are shown in Fig.
\ref{fig1}(c).

For the TDO measurement, a sample with dimensions $1.7\times
1.1\times 3.1$ mm$^3$ was placed inside a solenoid which is a part
of a {\it LC} tank circuit. The filling factor of the solenoid is
about 0.17 and the solenoid and sample was oriented such that
conducting planes are perpendicular to the external magnetic field
and parallel to the solenoid {\it rf} field. The detailed
experimental procedure is published elsewhere.\cite{brookstdo}

\section{Data and Results}
\subsection {Quantum Oscillations}
\begin{figure}[bp]
\epsfig{file=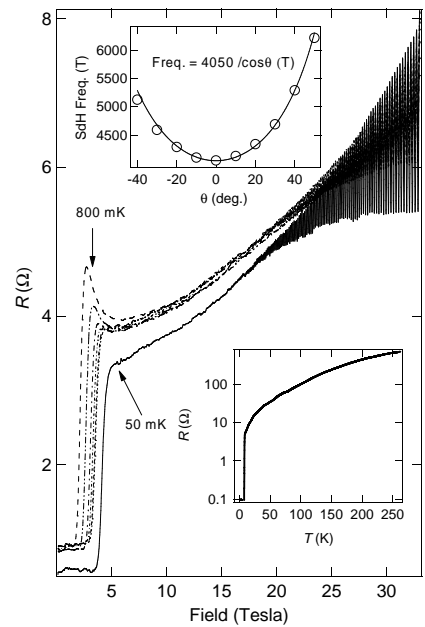,bb=71 94 530 720,width=0.7\linewidth}
\caption{Magnetoresistance of \sbf~ at different temperatures.
Insets : Resistance as a function of temperature at zero field and
angular dependence of SdH oscillation frequency.}\label{fig2}
\end{figure}

Fig. \ref{fig2} shows the magnetoresistance (MR) data at different
temperatures between 50 and 800 mK for the magnetic field
direction perpendicular to the conducting planes. Resistance as a
function of temperature at zero field, shown in the lower inset,
is different from the result of Ref. \onlinecite{yamada} who
reported a negative temperature coefficient of resistance (TCR) at
temperatures above about 180 K. However more recent
measurements\cite{shimojo} are consistent with our observation of
a positive TCR from room temperature to $T_c$. Although small
offsets are sometimes observed in the resistance below $T_c$, the
superconducting critical field transitions and quantum
oscillations are clearly observed in MR data. $H_{c2}$ shifts to
higher field with decreasing temperatures, and an anomalous peak
in MR can be seen near $H_{c2}$. These "hump" structures, commonly
observed in some organic superconductors, will be discussed below.
Only one frequency ($F_{\alpha}$) at about 4050 tesla was obtained
from the Fast Fourier Transform (FFT) as shown in the inset. The
angular dependence of this frequency follows a $1/\cos\theta$
dependence, which is expected for a two dimensional (2D),
cylindrical Fermi surface. The same oscillation frequency was
observed in the TDO measurements, also discussed below. Neither
beating behavior nor oscillations with different frequencies were
observed for the angles measured.

$F_{\alpha}$ corresponds to a Fermi surface with extremal area of
3.86$\times$10$^{15}$cm$^{-2}$, which is in good agreement with
the band calculation result. The effective cyclotron mass
($m_c^*$) and the Dingle temperature ($T_D$) can be obtained by
fitting the temperature and field dependence of the oscillation
amplitude to the Lifshitz-Kosevich (LK) formula.\cite{LK}
According to the 2D LK formula, the amplitude ($A$) of fundamental
oscillations can be expressed as,
\begin{alignat}{3}
\label{LK}
  A&=\frac {\alpha m_c^* T /B}{\sinh (\alpha m_c^* T /B)} \times \exp
(-\alpha m_c^* T_D /B) \times \cos (\frac {1}{2} \pi g m_b)\nonumber\\\
&=R_T \times R_D \times R_S,
\end{alignat}
where $\alpha$=14.69 T/K, $m_c^*$ is the effective cyclotron mass,
$T_D$ is the Dingle temperature, $g$ is the spin $g$-factor and
$m_b$ is the band effective mass. Each terms are associated with
damping effects due to finite temperature ($R_T$), electron
scattering ($R_D$) and spin splitting ($R_S$) respectively. (When
the oscillatory part of the magnetoresistance is used to determine
the effective mass from $R_T$, it is necessary to normalize this
signal by the background magnetoresistance, as we have done in the
following analysis.

\begin{figure}[tp]
\epsfig{file=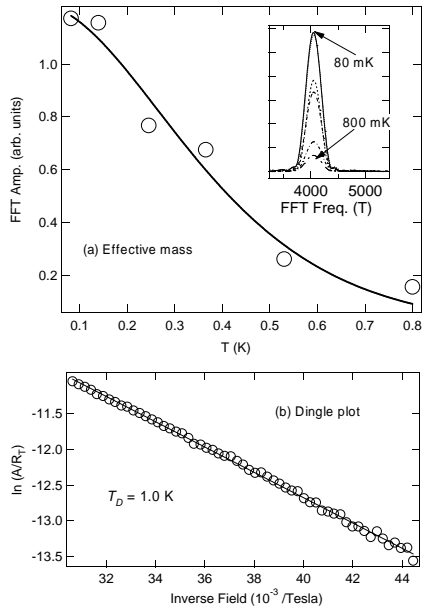,bb=60 60 535 730, clip=,width=0.7\linewidth}
\caption{(a) Effective mass fitting using the LK formula. The
inset shows FFT amplitudes for different temperatures used for the
effective mass fitting. (b) Dingle plot of \sbf} \label{fig3}
\end{figure}

The results of the LK fit are shown in Fig. \ref{fig3}. $m_c^*$
was estimated to be about 12.4 $\pm$ 1.1 $m_e$, which is quite
large compared to other organic conductors. $T_D$ is about 1.0 K,
which corresponds to a relaxation time $\tau \approx 1.1\times
10^{-12}$ sec and a mean free path $l \approx 795$ \AA. Due to
large $m_c^*$, the condition $\omega_c \tau > 1$ is satisfied when
the magnetic field is larger than 64 T, which is probably why
quantum oscillations were not observed at lower fields in previous
work.

\begin{figure}[tp]
\epsfig{file=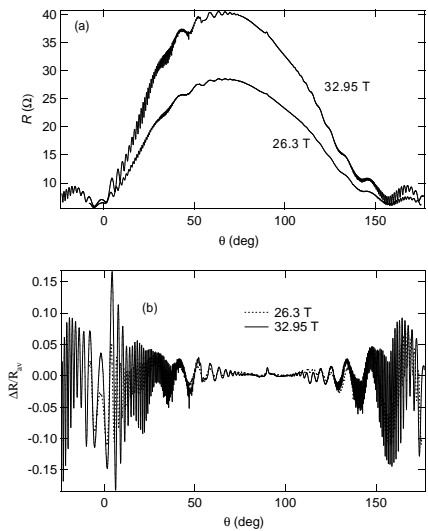,bb=50 120 540 720,
clip=,width=0.7\linewidth} \caption{(a) AMRO data at 50 mK for two
different fields and (b) AMRO amplitude after subtracting
background magnetoresistance.} \label{fig4}
\end{figure}

\begin{figure}[tp]
\epsfig{file=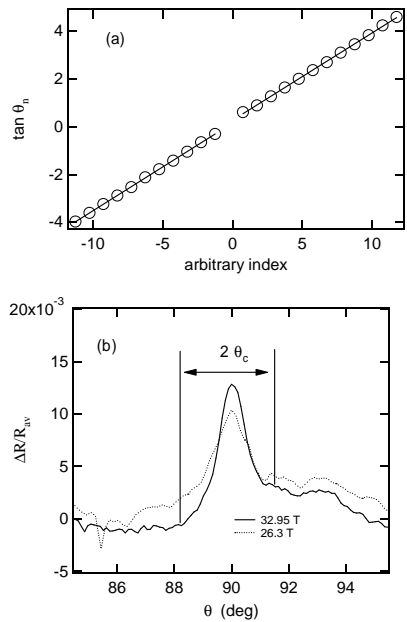,bb=90 150 480 780,
clip=,width=0.7\linewidth} \caption{(a) $\tan\theta$ plot for
Yamaji oscillations. (b) Expansion of AMRO amplitude near
$\theta$=90$^\circ$.} \label{fig5}
\end{figure}

Fig. \ref{fig4}(a) shows the AMRO data of \sbf~ at 50 mK when the
magnetic field is fixed at 32.95 and 26.3 tesla. The AMRO signal
can be more clearly seen by subtracting the background MR as shown
in Fig. \ref{fig4}(b). Besides the SdH oscillations for $\theta$
near 0 and 180$^\circ$, additional oscillations, so called Yamaji
oscillations,\cite{yamaji} are clearly seen for 20$^\circ$ $\leq$
$\theta$ $\leq$ 80$^\circ$ and 100$^\circ$ $\leq$ $\theta$ $\leq$
160$^\circ$. The peaks in the Yamaji oscillation are periodic in
$\tan\theta$ as shown in Fig. \ref{fig5}(a), which indicates that
the cylindrical Fermi surface is corrugated along the
$b^*$-direction. The periodicity of $\tan\theta$ is related to the
maximum Fermi wave vector projection to the field rotation plane
$k_F$ by the relation $bk_F \tan\theta=\pi(n-1/4)$, where $b$ is
the lattice parameter along the $b$-axis. From the slope of Fig.
\ref{fig5}(a), $k_F$ is estimated to be about 4.89$\times10^7$
cm$^{-1}$, which gives cross-sectional area $\pi k_{F}^2 =
7.5\times10^{15}$ cm$^{-2}$, two times larger than the area from
the SdH oscillation. Furthermore, the $k_F$ values is comparable
to the distance of $\Gamma$Z in the momentum space. The
discrepancy of Fermi surface areas between two experimental values
obtained from AMRO and SdH effect is not unusual and it can be
ascribed to higher order terms of Yamaji oscillations\cite{uji} or
effective doubling of the interlayer distance\cite{mielke}.

In Fig. \ref{fig5}(b), a peak in resistance (see also Fig. 4) near
$\theta$=$90^\circ$ is clearly shown. A peak in AMRO for the
 magnetic field parallel to the conducting layers has been observed in many organic
conductors, and is used as a proof of coherent interlayer
transport.\cite{hanasaki} Other experimental results, like a beat
frequency in the magnetic oscillations and a crossover from a
linear to a quadratic field dependence have also been suggested as
a test for coherent transport.\cite{moses} However, no beating
behavior was observed for our data, which may be ascribed to very
small value of interlayer transfer integral ($t_b$) compared to
the Fermi energy ($\epsilon_F$). The ratio of the interlayer and
the Fermi energy ($t_b / \epsilon_F$) can be estimated from the
relation $t_b / \epsilon_F = (\pi/2 - \theta_c )/bk_F $, where
$\theta_c$ is the critical angle above which the resistance
increases rapidly.\cite{hanasaki} Using the crystallographic
parameters and the experimental value of $k_F$ obtained from the
SdH oscillation, $t_b/\epsilon_F$ is estimated to be about 1/250,
and 1/350 when $k_F$ from AMRO is used. This value is comparable
to \betah~(1/175) and \ibr~(1/280) which both show beating
behavior\cite{wosnitza1}. Hence the reason why we do not observe a
beat frequency in \sbf~is not clear yet.

\begin{figure}[bp]
\epsfig{file=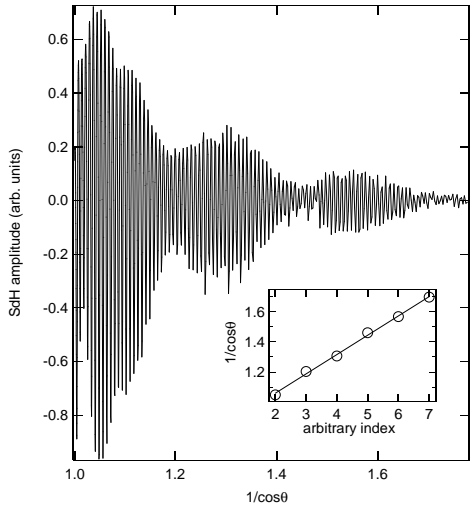,bb=60 216 528 714,
clip=,width=0.75\linewidth} \caption{(a) SdH oscillation amplitude
after subtracting AMRO from the data of Fig. \ref{fig4} as a
function of $\cos\theta$. Inset : Positions of nodes and antinodes
of SdH amplitude in 1$/\cos\theta$} \label{fig6}
\end{figure}

The amplitude of SdH oscillations in AMRO data can be obtained by
subtracting AMRO and background magnetoresistance. The result is
shown in Fig. \ref{fig6} for $H$=32.95 tesla. There are some nodes
periodic in $1/\cos\theta$, which can be more clearly seen by
plotting the locations of nodes and anti-nodes in $1/\cos\theta$
against odd and even integers. This behavior in AMRO can be
explained by the spin splitting term $R_S$ in Eq. \ref{LK}, where
$\cos(\pi gm_b /2)$ will show minima (maxima) when $gm_b$ are odd
(even) integers. Since the magnetic field is constant in AMRO
measurements, the nodes come from the angular dependence of
$m_b$($\theta$). In the framework of Fermi liquid theory, $m_c^*$
can be written as $m_c^*$=$m_b$(1+$\lambda$), where $\lambda$ is
the mass enhancement factor. And the angular dependence of $m_c^*$
is known to follow $1/\cos\theta$ dependence. From the slope of
$1/\cos\theta$ plot, we obtained $gm_b \sim$ 15.6$m_e$, which is
about 30\% larger than the largest known value in pervious organic
conductors.\cite{wosbook} The $gm_b$ values for representative
organic conductors are listed in Table \ref{table1}.

\subsection {Superconductivity}
\begin{figure}[bp]
\epsfig{file=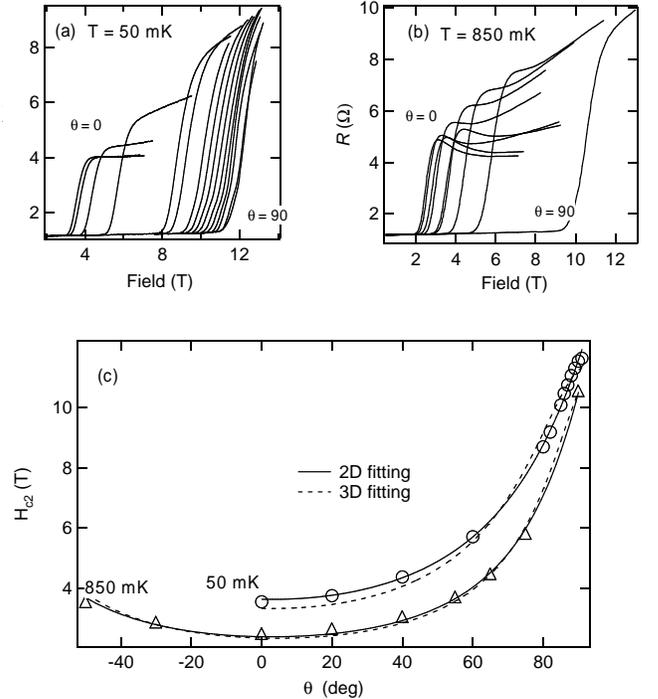,bb=53 160 561 730, clip=,width=\linewidth}
\caption{(a) Magnetoresistance near $H_{c2}$ at different angles
when $T$= $T$=50 mK (a) and 850 mK (b) $H{c2}$ as a function of
$\theta$. The fitted curves for the 2D and 3D model are also
shown.} \label{fig7}
\end{figure}

Fig. \ref{fig7} shows the angular dependence of $H_{c2}$($\theta$)
at two different temperatures. $H_{c2}$ was determined by the
field where the first derivative of $R(H)$ shows an extremum. The
values at $\theta$=0 and 90$^\circ$ are in good agreement with the
previous report when the $H_{c2}$($T$) curves in Ref.
\onlinecite{shimojo} are extrapolated to lower temperatures. The
solid and dashed line in Fig. \ref{fig7}(c)are the fitting curves
for the case of a 2D film\cite{tinkham} and using the 3D
anisotropic GL theroy\cite{moris}, respectively. Each can be
expressed as
\begin{equation}
\label{2d}
1=\left|\frac{H_{c2}(\theta)\cos\theta}{H_{c2\perp}}\right|+\left(\frac{H_{c2}(\theta)\sin\theta}{H_{c2\parallel}}\right)^{2}~~:\mathrm{2D},
\end{equation}
\begin{equation}
\label{3D} H_{c2}(\theta)=\frac{\phi_0}{2\pi\xi_\parallel^2
\sqrt{\cos^2 \theta+\frac{H_{c2\perp}^2}{H_{c2\parallel}^2}\sin^2
\theta}}~~:\mathrm{3D},
\end{equation}
where $\phi$ is the flux quantum and $\xi$ is the coherence
length. The validity of the application of these models depends on
relative magnitude of the inter-layer coherence length
($\xi_\perp$) and inter-layer distance ($b$), i.e., the 2D model
is more appropriate for $\xi_\perp < b$ and vice versa. Although
both fitting curves give approximate fits to the data, the 2D
model gives slightly better results. For comparison,
\kappai~superconductors show quite satisfactory fitting results
using the 2D model, and their quantum oscillation also show
peculiar features which were ascribed to nearly perfect
two-dimensionality.\cite{wosbook,wankaKI3} If one considers only
the criterion of the two lengths, which are $\xi_\perp$ (26 \AA)
and $b$ (17.6 \AA), the 3D model seems to be more appropriate.
Nevertheless, $H_{c2}$($\theta$) is described better by the 2D
formula.

The anomalous behavior in MR, a "hump" structure, near $H_{c2}$ is
also seen in Fig. \ref{fig7}(b). The hump structure is most
significant for $\theta$=0 ($H \perp$ conducting planes) and
almost disappears for $\theta=90^\circ$ ($H
\parallel$ conducting planes). This behavior was observed for
other $\kappa$- and $\beta^{\prime\prime}$-type organic
superconductors, which share several common features.\cite{zuo} It
is highly anisotropic for both electric current and magnetic field
directions; it is most significant when the inter-plane resistance
is measured with the magnetic field perpendicular to the
conducting planes. And it is suppressed with decreasing
temperature and increasing uniaxial stress along the inter-plane
direction.\cite{choi} In \sbf, the temperature dependence (see
Fig. \ref{fig2}) and anisotropic features are also apparent, while
the hump structure in this material can be still seen at very low
temperatures down to 80 mK.

\subsection {Tunnel Diode Oscillator Study}
\begin{figure}[bp]
\epsfig{file=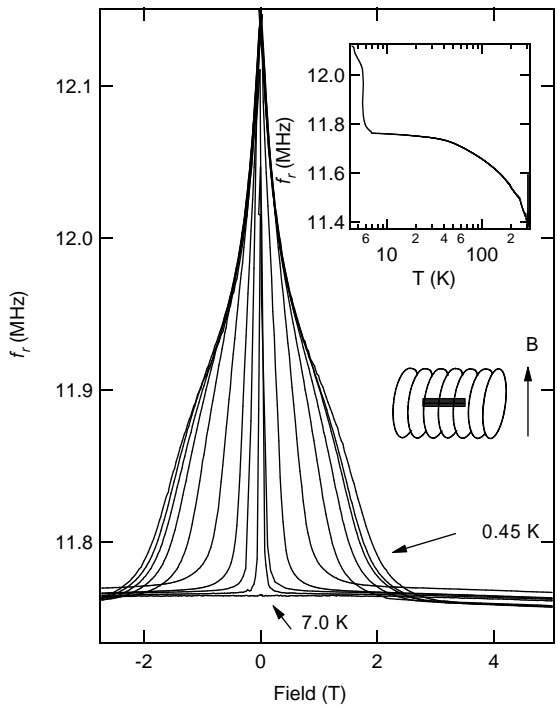,bb=64 168 529 695, clip=,width=\linewidth}
\caption{TDO resonant frequency ($f_r$) at different temperatures.
The inset shows the temperature dependence of $f_r$ at zero
field.}\label{fig8}
\end{figure}

Fig. \ref{fig8} shows the field dependence of a resonant frequency
($f_r$) of TDO at low field regions between 7.0 and 0.45 K. When
the temperature is below $T_c$, the frequency decreases sharply
with increasing field until superconducting-normal transition
occurs. In the normal state, the frequency change is less
significant showing gradual decrease. The difference in magnitude
of the resonant frequency in the superconducting and normal state
can be also seen in the zero-field cool-down curve as shown in the
inset of Fig. \ref{fig8}. Since the resonant frequency of an {\it
LC} circuit is $f_r$=$1/(2\pi\sqrt{LC})$, the change of $f_r$
mainly comes from the change of inductance of the coil.

The decrease of $f_r$, i.e., the increase of $L$ in
superconducting-metal transitions can be ascribed to the
penetration of the {\it rf} field to the sample.\cite{mielke,wu}
The {\it rf} field will be expelled when a sample is in the
superconducting state, which results in a decrease of $L$ and
increase of $f_r$. Following this argument, a $H$-$T$ phase
diagram can be built based on the TDO experiment. $H_{c2}$ was
defined as a intersection point of extrapolations of $f_r (H)$
above and below the transition. And the field where $f_r (H)$ show
change of curvatures ($H^*$) was defined as a point that $f_r (H)$
starts to deviate $H^{1/2}$ dependence. Our work shows similar
behavior compared with the results of Ref. \onlinecite{mielke}, in
which $H^*$ in  \gacl4~superconductor was ascribed to the change
of fluxoid motion in the mixed state. According to this model,
fluxoid motion can be explained by a simple harmonic oscillator
when $H \ll H_{c2}$ and $H<H^*$. In this regime, the frequency
change, or equivalently the penetration depth follows $H^{1/2}$
dependence. Above $H^*$, the deviation from the $H^{1/2}$ behavior
was attributed to the flux-lattice melting, which occurs when the
thermal displacement of vortex lattice becomes significant
compared to lattice parameters.\cite{houghton}

\begin{figure}[tp]
\epsfig{file=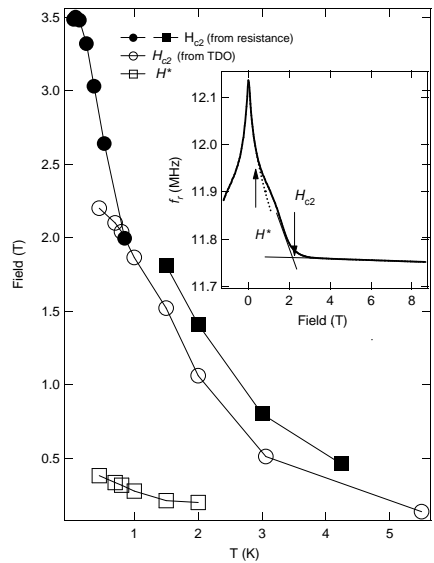,bb=61 110 535 712,
clip=,width=0.7\linewidth} \caption{$T$-$H$ phase diagram based on
TDO measurements and resistance measurements done two different
samples. Inset : definition of $H_{c2}$ and $H^*$ in the TDO
experiment}\label{fig9}
\end{figure}

The resulting $H$-$T$ phase diagram is shown in Fig. \ref{fig9},
where $H_{c2}$ values from the resistance measurement on two
samples are also plotted. $H_{c2}$ values from the two different
experiments overlap at $T\approx$ 0.8 K, but $H_{c2}$ values from
the resistance measurement seem to show more pronounced
temperature dependence at lower temperatures. The $H_{c2}(T)$
curves do not follow the power law $H_{c2}\propto(T_c -T)^n$
behavior in the whole temperature range, rather there seems to be
a change of curvature around $T$=2 K.

\begin{figure}[tp]
\epsfig{file=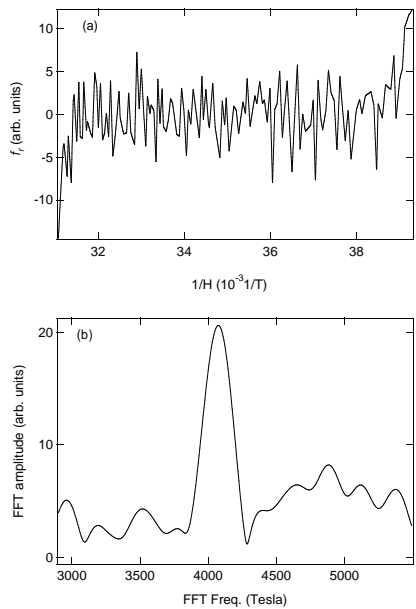,bb=56 58 536 727,
clip=,width=0.7\linewidth} \caption{TDO resonant frequency at high
field region for $T$=0.45 K} \label{fig10}
\end{figure}

The amplitudes of quantum oscillations of $f_r$ in the normal
state between 25 and 32 tesla is shown as a function of inverse
field in Fig. \ref{fig10}. The overall signal to noise ratio is so
small that it is difficult to resolve the nature of quantum
oscillations directly from the TDO data. However, the oscillation
frequency obtained from FFT is about 4070 T, very similar to that
of the resistance measurement.

\section {Discussion} The electronic structure and
superconductivity of \sbf~have many common features with other
organic superconductors. The molecular packing mode of this
material is known to be very similar to that of \betai.
Correspondingly, the Fermi surface of \sbf~resembles the other
$\beta$-type organic conductors both in the band calculation and
experimental results. The superconductivity in \sbf~ also has
similar characteristics to other ET and BETS organic
superconductors. $H_{c2}(\theta)$ measurements revealed the
anisotropic nature of the superconductivity in \sbf. There could
be a possibility of a dimensional crossover due to the increase of
the perpendicular coherence length $\xi_{\perp}$ at higher
temperatures than $H_{c2}(\theta)$ measurements were done in this
work.\cite{wankaKI3} The deviation from the power law of
$H_{c2}(T)$ curves could be a indication of a dimensional
crossover, which was suggested for \gacl4~
superconductors.\cite{mielke}

In Table \ref{table1}, we have listed some physical parameters
related to fermiology and superconductivity of several organic
superconductors including \sbf. $F_\alpha$ and $m_{c\alpha}^*$
denote the FFT frequency and effective cyclotron mass for
fundamental orbits and $F_\beta$ and $m_{c\beta}^*$ for breakdown
orbits. When two orbits coexist, the value for the $\beta$-orbit
is more appropriate for comparison, since it covers a larger area
of the Fermi surface.

\begin{table*}
\caption{A comparison for fermiological and superconducting
properties or organic superconductors\label{table1}}
\begin{ruledtabular}
\begin{tabular}{llllllllll}
&$F_\alpha$&$m_{c\alpha}^*$&$F_\beta$&$m_{c\beta}^*$&$gm_b$&$\gamma$&$\Delta C_e/\gamma T_c$&$T_c$\\
&(T)&($m_e$)&(T)&($m_e$)&&(mJ/mol$\cdot$K$^2$)&&(K)\\ \colrule
\\
\sbf&4050&12.4&&&15.3&33\cite{shimojo}&1.1\cite{shimojo}&6.5\\
\cunbr\cite{elsinger,weis}&530&3.1&3790&6.6&9.85&25&2.7&11.5\\
\cuncs\cite{caulfield,andraka}&600&3.5&3920&6.5&5.2&25&2.0-2.8&10.4\\
\betah\cite{kang,wosbook}&3730&4.2&&&11.96&&&8.1\\
\sf5\cite{wosphysicaB,wankacp}&200&1.9&&&4&18.7&2.1&5.2\\
\gacl4\cite{mielke}&650&3.6&4030&6.3&&&&5\\
\kappai\cite{wosbook,wosnitzaCp}&570&1.9&3880&3.9&8.64&18.9&1.6&3.6\\
\ibr\cite{kartsovnik,wosbook}&3900&5.0&&&8.99&&&2.7\\
\nh4\cite{wosbook,wosnh4}&560&2.7&&&4.4&25&1.0&1.2\\
\end{tabular}
\end{ruledtabular}
\end{table*}

The striking difference between \sbf~and other organic
superconductors can be found in values of $m_c^*$ and $gm_b$.
Since the Fermi surface of \sbf~ is similar to other $\beta$-type
organic conductors, the difference can be attributed to many body
effects which are not considered in the band calculation.

In the framework of Fermi liquid theory, several effective masses
can be defined.\cite{merino} Experimentally, SdH and dHvA effect
can be used to obtain the effective cyclotron mass $m_c^*$ which
reflects mass enhancement by both electron-phonon and
electron-electron interactions. The dynamic mass $m_c$ is
influenced only by electron-phonon interactions and can be
measured by the cyclotron resonance experiment. The band mass
$m_b$ has similar value to the optical mass which can be obtained
by the plasma frequency measurement. The Sommerfeld constant
$\gamma$ which can be measured from the specific heat measurement
is related to the specific heat mass that is affected by both
interactions. The mass enhancement of $m_c^*$ due to
electron-phonon and electron-electron interactions can be written
as,
\begin{equation}
\label{mass} m_c^*=(1+\lambda_{EP})(1+\lambda_{EE}) m_b =
(1+\lambda_{EE}) m_c,
\end{equation}
where $\lambda_{EP}$ and $\lambda_{EE}$ are the coupling strength
of electron-phonon and electron-electron interactions
respectively.

For several organic superconductors, \lph~obtained from the
magnitude of \delc~shows general tendency with respect to $T_c$,
which is in agreement with weak and strong coupling limit of BCS
theory.\cite{elsinger,wankacp} But \sbf~does not seem to agree
with this tendency considering measured value of \delc~and $T_c$.
The small value of \delc~of \sbf~suggests that it is below the
weak coupling limit like \nh4~, but its $T_c$ value is between
\sf5~and \cuncs~which are in the range of intermediate and strong
coupling limit (see Fig. 3 in Ref. \onlinecite{elsinger} and Table
\ref{table1}). This apparent discrepancy suggest that additional
parameters should be considered for \sbf~ superconductor.

Considering the large value of $m_c^*$ of \sbf~ which cannot be
explained by \lph~alone, it is likely that electron-electron
interactions play an important role in this material. The large
values of $gm_b$ and $\gamma$ are also suggestive that the
effective density of states (quasiparticle density of states) is
large compared to other organic conductors. Indeed, the largest
transfer integrals obtained from the band calculation were about
half of those of \cuncs~and \betai,\cite{yamada} which suggest
that the narrower bandwidth could be the reason for the enhanced
electron-electron interactions. At present, it is difficult to
estimate the magnitudes of $\lambda_{EP}$ and $\lambda_{EE}$ until
pressure and cyclotron resonance studies are carried out.

In conclusion, the new donor molecule organic superconductor
\sbf~is found to have similar properties in the electronic
structure and superconductivity to other typical organic
superconductors. But the large effective mass indicates that many
body effects are significant in this material. In the context,
this new type of donor molecules could be a new approach to
increase $T_c$ of organic conductors by enhanced electron-electron
interactions.

\begin{acknowledgments}
This work was supported by NSF-DMR 99-71474. The National High
Magnetic Field Laboratory is supported by a contractual agreement
between the National Science Foundation and the State of Florida.
A. Wade and E. Goetz were supported by REU fellowships through the
Department of Physics (FSU) and the National Science
Foundation(NHMFL) respectively.
\end{acknowledgments}



\newpage

\end{document}